\documentclass{iau} 
\usepackage{graphicx}

\def\kms{km s$^{-1}$}
\newcommand{\msun}{$M_\odot$}

\title[High-mass Star Formation in G22] 
{Filamentary Flows and Clump-fed High-mass Star Formation in G22}

\author[J. Yuan et al.]   
{J. Yuan$^1$, J.-Z. Li$^1$, Y. Wu $^2$}

\affiliation{$^1$National Astronomical Observatories, 
	Chinese Academy of Sciences, 
	20A Datun Road, Chaoyang District, Beijing 100012, 
	China; email: {\tt jhyuan@nao.cas.cn} \\[\affilskip]
{$^2$Department of Astronomy, Peking University, 
	100871 Beijing, China;}}

\pubyear{2017}
\volume{336}  
\jname{Astrophysical Masers: Unlocking the Mysteries of the Universe}
\editors{A. Tarchi, M.J. Reid \& P. Castangia, eds.}
\begin{document}

\maketitle

\begin{abstract}
G22 is a hub-filament system composed of four 
supercritical filaments. Velocity gradients are detected along 
three filaments. A total mass infall rate of 440 
\msun~Myr$^{-1}$ would double the hub mass in about six 
free-fall times. The most massive clump C1 would be 
in global collapse with an 
infall velocity of 0.31 \kms~and a mass infall rate of 
$ 7.2\times10^{-4} $ \msun~yr$^{-1}$, which is supported by 
the prevalent HCO$^+$ (3-2) and $^{13}$CO (3-2) blue profiles. 
A hot molecular core (SMA1) was revealed in C1. 
At the SMA1 center, there is a massive 
protostar (MIR1) driving multipolar outflows which are associated 
with clusters of class I methanol masers. 
MIR1 may be still growing with an accretion rate of 
$7\times10^{-5}$ \msun~yr$^{-1}$. Filamentary flows, 
clump-scale collapse, core-scale accretion coexist in G22, 
suggesting that  
high-mass starless cores may not be prerequisite to form 
high-mass stars. In the high-mass star formation process, 
the central protostar, the core, and the clump can grow in mass 
simultaneously.
\end{abstract}

\firstsection 
\section{Introduction}

The two extensive debated high-mass star formation 
scenarios have plotted largely different pictures of
mass accumulation process. 
Extensive investigations show that the prevalent filaments
are the most important engine of forming stars, especially 
for the high-mass ones (\cite[Andr\'{e} et al. 2014]{Andre14}).
How the gas flows detected in filaments help 
individual cores grow in mass is still a key open question.
In this work, a filamentary cloud G22 is extensively 
investigated to reveal a promising mass accumulation
scenario.

\section{Mass accumulation process in G22}
\begin{figure}
	\begin{center}
		\includegraphics[width=0.9\textwidth]{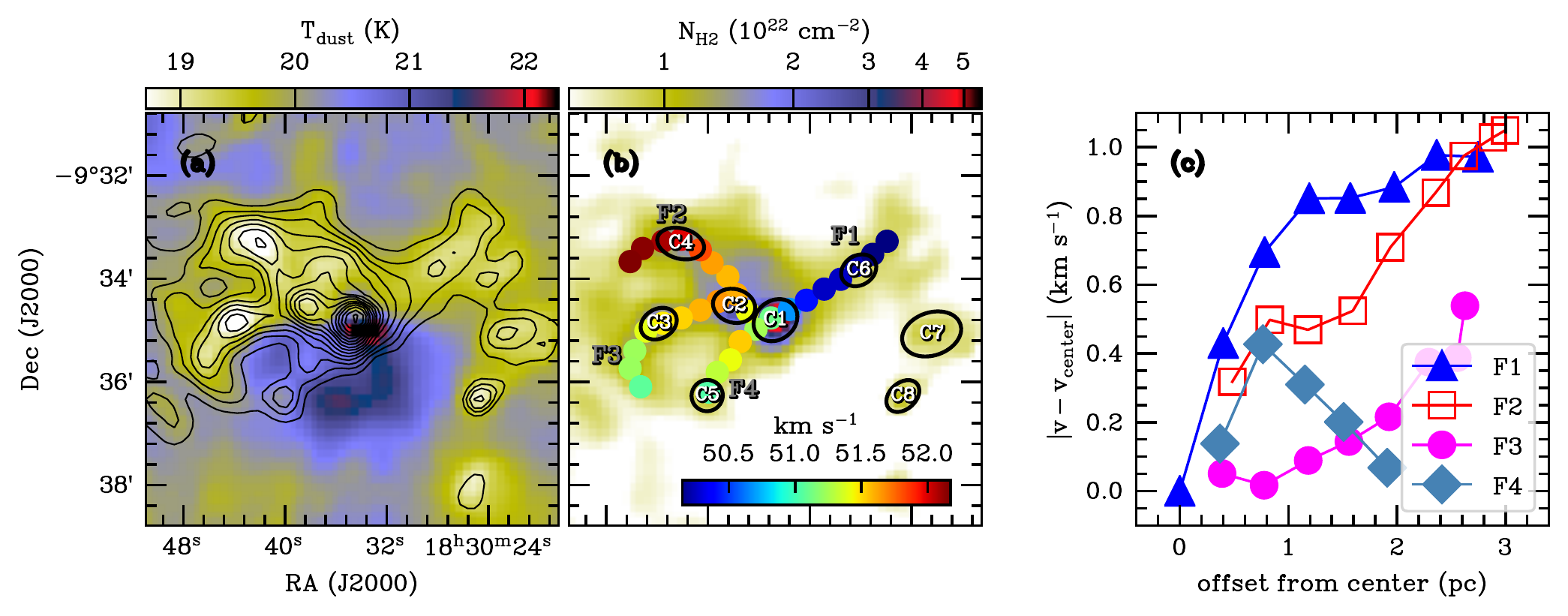} 
		\caption{(a) Dust temperature map from SED 
			fitting with column density overlaid as contours. 
			(b) Velocity centroids of $^{13}$CO (1-0) 
			spectra extracted along filaments overlaid 
			on top of the $N_\mathrm{H_2}$ map. The eight clumps, 
			designated as C1 to C8, are shown as open 
			ellipses. (c) LOS velocity of $^{13}$CO as a 
			function of distance from the potential well 
			centers, i.e., clump C1 for F1, F2, and F4, 
			and clump C2 for F3}
		\label{fig1}
	\end{center}
\end{figure}

\begin{figure}
	\begin{center}
		\includegraphics[width=0.9\textwidth]{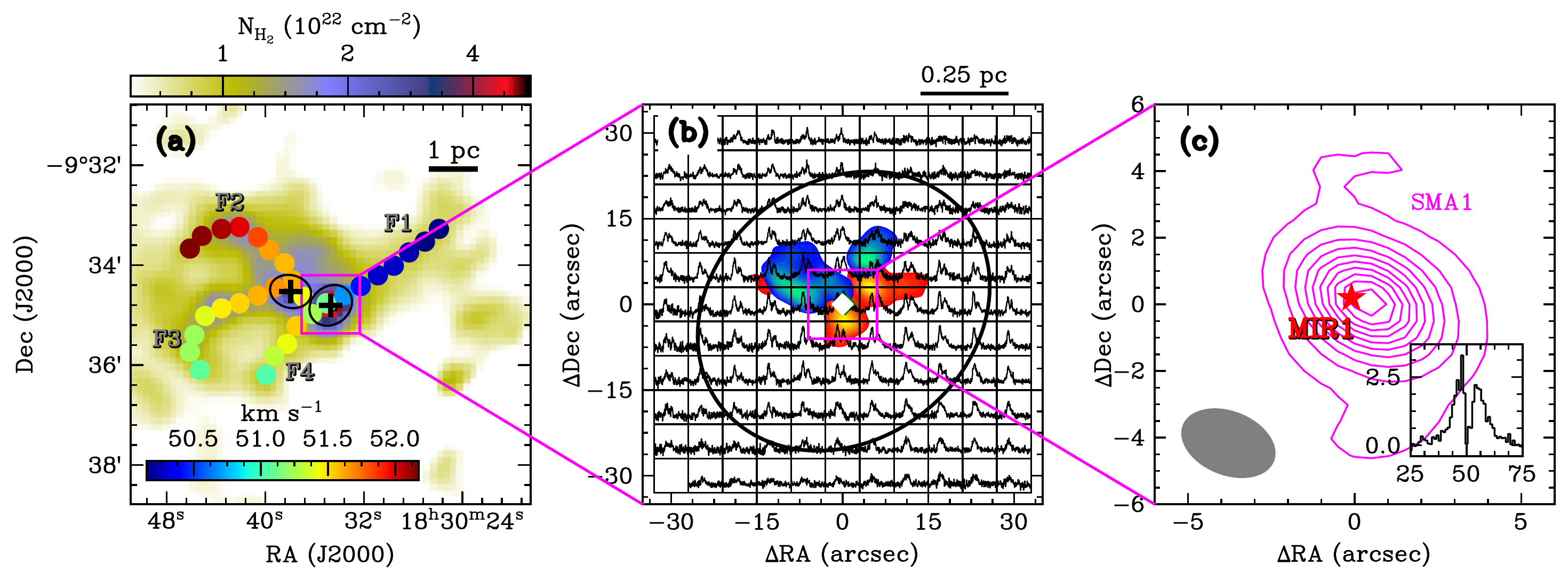} 
		\caption{(a) Velocity centroids of $^{13}$CO (1-0) 
			extracted along filaments overlaid on 
			the $N_\mathrm{H_2}$ map. 
			(b) Spectra of JCMT/ $^{13}$CO (3-2) 
			overlaid on SMA/CO (2-1) outflows. The 
			large ellipse delineates clump C1. 
			(c) A close-up view of the SMA 1.3 mm continuum. 
			The mono-core is designated as SMA1. A filled star 
			shows the MIR source SSTGLMC G022.0387+00.2222 
			(MIR1) from the GLIMPSE survey. The insert plot shows
		    the $^{13}$CO (2-1) spectrum at the SMA1 peak.}
		\label{fig2}
	\end{center}
\end{figure}

	{\underline{\it G22: a collapsing hub-filament system}}. 
	With a distance of 3.51 kpc, the G22 cloud contains ten 
	\textit{Spitzer} infrared dark clouds (IRDCs). These IRDCs are mainly 
	distributed in in a hub-filament system. 
	As shown in Figure \ref{fig1} (b), systematic velocity changes are
	detected along filaments F1, F1, and F3 based on $^{13}$CO (1-0)
	observations. The differences between the velocities of the filaments
	and the junction as a function distance to the center shows 
	monolithically increasing profiles for F1, F2, and F3
	(see Figure \ref{fig1} (c)). This
	suggests that gas is transfered to the hub region along these 
	filaments with an estimated total mass infall rate of 440 
	\msun~Myr$^{-1}$ (\cite[Yuan et al. 2017]{Yuan17}).
	
	{\underline{\it G22-C1: a collapsing high-mass clump}}. 
	Located at the hub region, C1 is the most massive
	clump with a mass of 590 \msun. Prevalent blue profiles
	are detected toward C1 (see Figure \ref{fig2} (b)), suggestive
	of clump-scale global collapse. The estimated mass infall rate is
	$ 7.2\times10^{-4} $ \msun~yr$^{-1}$
	
	{\underline{\it G22-C1-SMA1: A collapsing hot molecular core}}. 
	At the center of C1, a hot molecular core SMA1 with a gas temperature
	higher than 220 K is detected. The spectrum of $^{13}$CO (2-1) and 
	C$^{18}$O (2-1)
	show blue profiles (see Figure \ref{fig2} (c)), 
	indicating infall motions in SMA1. The estimated
	mass accretion rate is about $7\times10^{-5}$ \msun~yr$^{-1}$.

\section{Conclusions}
	
	Inward motions have been detected along filaments, in the center
	clump and dense core. The continuous mass growth from large
	to small scales suggests that high-mass starless cores might not 
	be prerequisite to form high-mass stars. The deeply embedded protostar,
	the core, and the clump can simultaneously grow in mass.

\end{document}